# Optimizing the electronic control loop of a solid-state ring laser gyroscope

S. Schwartz, M. Rebut, G. Feugnet, J. Colineau and J.-P. Pocholle

*Abstract :*

We study in this Letter the dynamical effects of the limited bandwidth of the control electronics in a solid-state (Nd-YAG) ring laser gyroscope. We derive a stability condition for the rotation-sensing regime in the case of a first-order control loop, showing that the smallest measurable rotation speed depends directly on the cutoff frequency value. Our experimental measurements are in good agreement with this prediction.

*Introduction*: The Ring Laser Gyroscope (RLG) is a high-precision rotation sensor used for inertial navigation [1-3]. It typically consists of a ring laser cavity emitting two counterpropagating beams. When such a device is rotated, the eigenfrequency difference $\Omega/(2\pi)$ between the counterpropagating waves is proportional to the speed of rotation $\dot{\theta}$ according to the following law :

$$\Omega = 2\pi\, S\dot{\theta}$$

where *S*, which depends on the design of the cavity, is known as the scale factor. The frequency of the beat note signal between the counterpropagating waves thus provides a convenient and accurate measurement of angular motions. In order to ensure bidirectional operation, which is required to obtain the beat signal, a gaseous amplifying medium is typically used, circumventing mode competition [4] thanks to Doppler gain broadening (see [1-3]). However, many industrial constraints arise from the use of a gas, which is usually the limiting factor in terms of lifetime, cost and reliability of the RLG. Another option is to use a solid-state gain medium, with an electronic feedback loop for intensity control and stabilization. This option has been previously described and achieved in [5,6], assuming an infinite bandwidth for the theoretical modelization of the electronic feedback loop. In this Letter, we study analytically and experimentally the effects of a limited bandwidth, and we show, in the case of a first order loop, that the value of the smallest measurable rotation speed (i.e. the performance of the gyroscope) directly depends on the loop cutoff frequency. Our experimental results, obtained with a diode-pumped Nd-YAG ring laser gyroscope, are in good agreement with our theoretical predictions. This work could find applications in the design of a new generation of competitive diode-pumped solid-state ring laser gyroscopes.

*Theoretical description*: The total electric field inside the ring laser cavity is assumed to be the sum of two counterpropagating modes :

$$\mathbf{E}(x,t) = \mathrm{Re}\left[\widetilde{E}_1(t)e^{i(\omega_c t - kx)} + \widetilde{E}_2(t)e^{i(\omega_c t + kx)}\right]\mathbf{e}$$

where $\widetilde{E}_1$ and $\widetilde{E}_2$ are the amplitudes of the modes, slowly-varying with respect to the mean eigenfrequency $\omega_c$, $k$ is the wavevector associated with the longitudinal coordinate *x* and **e** is a unitary vector defining the (common) state of polarization. The dynamics of the diode-pumped Nd-YAG ring laser is then ruled by the following equations [5,6] :

$$\frac{d\tilde{E}_{1,2}}{dt} = (-1)^{1,2}\left[\frac{\Delta}{2} + \frac{i\Omega}{2}\tilde{E}_{1,2}\right] - \frac{\gamma}{2}\tilde{E}_{1,2} + \frac{i\tilde{m}_{1,2}}{2}\tilde{E}_{2,1} + \frac{\sigma L}{2T}\left(\tilde{E}_{1,2}N_0 + \tilde{E}_{2,1}N_{1;2}\right) \quad (1)$$

where $\Delta$ is the difference between the loss coefficients of each mode, $\gamma$ is the average value of these loss coefficients, $\tilde{m}_{1,2}$ are phenomenological parameters describing backscattering on the cold cavity elements, $\sigma$ is the emission cross section, $L$ is the cavity length, $T$ is the cavity round trip time and $N_0$ and $N_{1,2}$ are defined by :

$$N_0 = \frac{1}{L}\int_0^L N dx \quad \text{and} \quad N_{1,2} = \frac{1}{L}\int_0^L Ne^{2i(-1)^{2,1}kx}dx,$$

$N$ being the population inversion density function, obeying the following equation :

$$\frac{\partial N}{\partial t} = \frac{W_{th}}{T_1}(1+\eta) - \frac{N}{T_1} - \frac{aN}{2T_1}\left|\tilde{E}_1 e^{-ikx} + \tilde{E}_2 e^{ikx}\right|^2$$

where $W_{th}$ is the pumping rate at laser threshold, $T_1$ is the population inversion lifetime (we assume $T_1$=230 µs in what follows), $\eta$ is the excess of pump power above threshold and $a$ is the saturation parameter. We consider the regime of beat note operation, i.e. the regime where the intensities of both modes are approximately constant and equal and where the (angular) frequency difference between both modes is close to the ideal (angular) frequency difference $\Omega$. In the high rotation rate limit, namely :

$$|\Omega| \gg |\tilde{m}_{1,2}| \quad \text{and} \quad |\Omega| \gg \sqrt{\gamma\eta/T_1} \quad (2)$$

the laser parameters for the beat regime have the following expressions (see reference [6]) :

$$|\tilde{E}_1|^2 + |\tilde{E}_2|^2 = B(t) + y_M(t) \quad \text{with} \quad |y_M| \ll B$$

$$\left|\tilde{E}_1\right|^2 - \left|\tilde{E}_2\right|^2 = C(t) + x_M(t) \quad \text{with} \quad |C|, |x_M| << B$$

$$\arg(\tilde{E}_2) - \arg(\tilde{E}_1) - \Omega t = \Phi_0(t) + \Phi_M(t) \quad \text{with} \quad |\Phi_0|, |\Phi_M| << 1$$

where the functions $x_M$, $y_M$ and $\Phi_M$ oscillate with an angular frequency close to $\Omega$ while $B$, $C$ and $\Phi_0$ are slowly varying with respect to the time scale $1/|\Omega|$. When the loss coefficients of the counterpropagating modes are equal (i.e. $\Delta = 0$), this regime is not a solution of the laser equations (1) and hence does not occur in practice [4-6]. To obtain the beat regime, one has to introduce an additional stabilizing coupling, which can be achieved for example by actively creating more losses for the strongest mode. This was modeled in [5,6] by assuming differential losses of the following form :

$$\Delta = aK\left(\left|\tilde{E}_1\right|^2 - \left|\tilde{E}_2\right|^2\right) \quad \text{with} \quad K > 0. \tag{3}$$

However, such a description does not take into account the way those differential losses are created experimentally, namely using an active electronic feedback loop acting on a Faraday magneto-optic element. An efficient yet simple way of designing the electronics configuration turns out to be the use of a first-order loop, resulting in differential losses $\Delta$ obeying the following equation :

$$\frac{d\Delta}{dt} = -\omega_0(\Delta - aKC) \tag{4}$$

where $\omega_0/(2\pi)$ is the cutoff frequency and where the effects of the intensity fast modulations $x_M(t)$ on the electronics response have been neglected. In the limit of an infinite cutoff frequency, equation (4) is consistent with the previous model (3). Equations (1), together with equation (4), can be solved following the procedure described in [6]. A stationary solution is found for the slowly varying quantities *B*, *C* and $\Delta$. Applying small perturbations like $e^{\lambda t}$ to this stationary solution leads to the following characteristic equation :

$$\lambda^2 + \lambda\left(\omega_0 - \frac{\gamma\eta}{2\Omega^2 T_1^2}\right) + \omega_0\left(K\eta - \frac{\gamma\eta}{2\Omega^2 T_1^2}\right) = 0. \tag{5}$$

The beat regime is stable when all the roots of equation (5) have a negative real part, which finally yields (using the Routh-Hurwitz criterion) the following simple conditions :

$$\omega_0 > \frac{\gamma\eta}{2\Omega^2 T_1^2} \qquad \text{and} \qquad K\eta > \frac{\gamma\eta}{2\Omega^2 T_1^2}. \tag{6}$$

In the infinite-bandwidth limit, conditions (6) reduce to the single condition on *K*, which is consistent with [5,6]. In the more realistic case of a limited bandwidth, both conditions (6) have to be fulfilled to ensure the stability of the beat regime.

*Experiment :* in order to confirm experimentally the beat regime stability conditions (6), we used a diode-pumped Nd-YAG ring laser gyroscope configuration similar to the one described in [5,6]. Concerning the electronic control loop, we ensured a first order behaviour, in particular thanks to a current controlled output stage connected to the solenoid of the magneto-optic element, which allows to cancel the frequency response effects of the inductive load. Typical parameters for our control loop are $\omega_0 \approx 10^4\,\text{s}^{-1}$ and $K \approx 10^7\,\text{s}^{-1}$, which implies in particular – see (6) – that the condition on $\omega_0$ is more stringent than the condition on *K*, and will ultimately determine the value of the lowest measurable rotation speed $\dot{\theta}_{cr}$, according to the following formula :

$$\dot{\theta}_{cr} = \frac{1}{2\pi S} \sqrt{\frac{\gamma \eta}{2\omega_0 T_1^2}} \,. \tag{7}$$

Considering equation (7), we have measured the value of $\dot{\theta}_{cr}$ as a function of the loop cutoff frequency $\omega_0/(2\pi)$. The result is reported on figure 1. Our experimental laser parameters are $\gamma = 24.1$ s$^{-1}$, $S=1.37$ kHz/(deg/s) and $\eta=0.8$. The precision in measuring $\dot{\theta}_{cr}$, typically ±5 deg/s, is mainly limited by the hysteretic behavior of the laser gyroscope at the boundary of the beat regime zone, due to the existence of metastable transient regimes with a frequency dependence very similar to the true stable beat regime. As can be seen on figure 1, the measured values for $\dot{\theta}_{cr}$ are in good agreement with our theoretical predictions (7) provided the cutoff frequency is not too high (typically below 500 Hz). For higher values of $\omega_0/(2\pi)$, the hypotheses (2) used in deriving (6) and (7) become more stringent than the second condition of (6) itself, which explains the fact that the agreement between equation (7) and the experimental data of figure 1 no longer stands. Instead, some other physical parameters such as the strength of the population inversion grating become determining in the laser gyroscope dynamics. We have also checked, considering equation (7), the dependence of $\dot{\theta}_{cr}$ on the pumping parameter $\eta$. For a cutoff frequency $\omega_0/(2\pi)=482$ Hz, we have measured $\dot{\theta}_{cr}=35$ deg/s when $\eta=0.8$, and $\dot{\theta}_{cr}=25$ deg/s when $\eta=0.4$. The ratio between those two measured values is very close to $\sqrt{2}$, which is consistent with equation (7).

*Conclusion* : the influence of the bandwidth of the electronic control loop on the solid-state ring laser gyroscope dynamics has been studied in the case of a first-order configuration. It is found theoretically that the performance of the gyroscope directly depends on the cutoff frequency provided this latter is not too high. Experimental data show very good agreement with this prediction. This result could find applications in the design of future solid-state high-performance inertial navigation devices.

*Acknowledgements :* the authors acknowledge constant support from Thales Aerospace Division. They also wish to thank Augustin Mignot and Mehdi Alouini for fruitful discussions.

**Authors' affiliations:**
S. Schwartz, M. Rebut, G. Feugnet, J. Colineau and J.-P. Pocholle (Thales Research and Technology France, RD 128, 91767 Palaiseau cedex, France)

e-mail address : sylvain.schwartz@thalesgroup.com

**Figure captions:**

Fig. 1 Smallest measurable rotation speed $\dot{\theta}_{cr}$ as a function of the control loop cutoff frequency $\omega_0/(2\pi)$ (linear vs logarithmic scale).

Figure 1

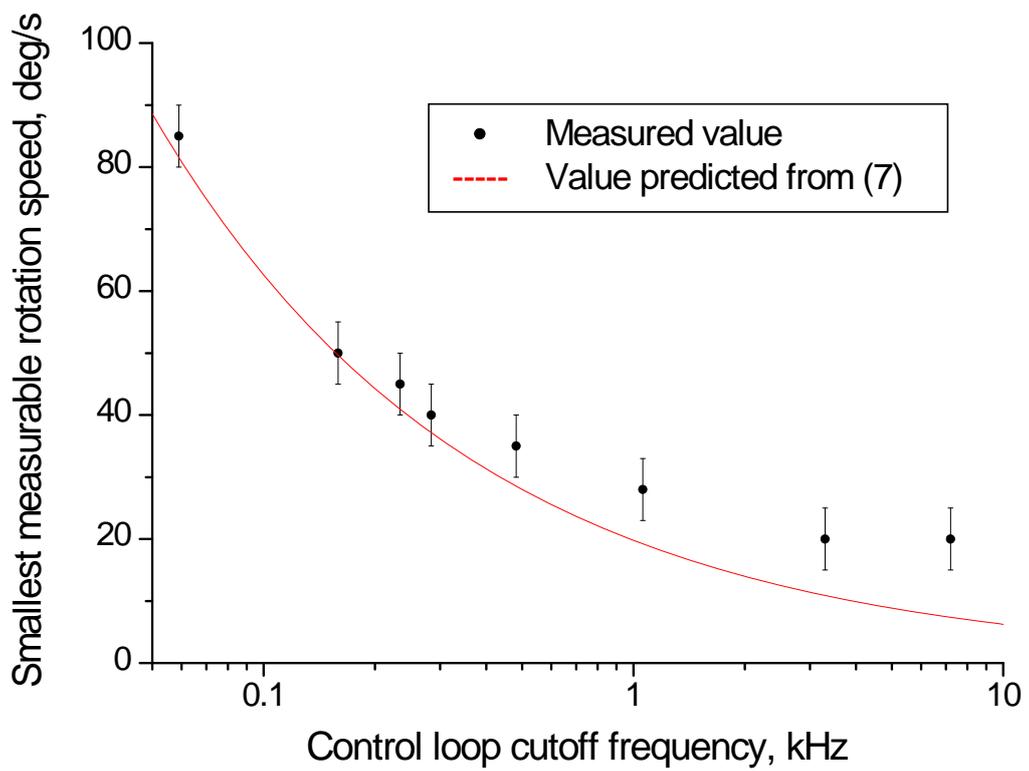